\documentclass[journal]{IEEEtran}
\ifCLASSINFOpdf
\else
\fi
\usepackage{mathtools}
\usepackage{color}
\usepackage{algorithm}
\usepackage[noend]{algpseudocode}

\usepackage{times}
\usepackage{epsfig}
\usepackage{graphicx}
\usepackage{amssymb}
\usepackage{romannum}
\usepackage{color}

\usepackage{amsmath}
\usepackage{breqn}
\usepackage{multirow}

\begin{document}
%
\title{Cell Type Identification from Single-Cell Transcriptomic Data via Semi-supervised Learning}
%
%
%

 \author{Xishuang Dong,~Shanta Chowdhury,~Uboho Victor,~Xiangfang Li, and~Lijun Qian
\thanks{X. Dong, S. Chowdhury, U. Victor, X. Li and L. Qian are with the Center of Excellence in Research and Education for Big Military Data Intelligence (CREDIT Center) and Center for Computational Systems Biology (CCSB), Department of Electrical and Computer Engineering, Prairie View A\&M University, Texas A\&M University System, Prairie View, TX 77446, USA. Email: xidong@pvamu.edu, shanta.chy10@gmail.com, uboho.dpc@outlook.com, xili@pvamu.edu, liqian@pvamu.edu}
}

\maketitle

\begin{abstract}
Cell type identification from single-cell transcriptomic data is a common goal of single-cell RNA sequencing (scRNAseq) data analysis. Neural networks have been employed to identify cell types from scRNAseq data with high performance. However, it requires a large mount of individual cells with accurate and unbiased annotated types to build the identification models. Unfortunately, labeling the scRNAseq data is cumbersome and time-consuming as it involves manual inspection of marker genes. To overcome this challenge, we propose a semi-supervised learning model to use unlabeled scRNAseq cells and limited amount of labeled scRNAseq cells to implement cell identification. Firstly, we transform the scRNAseq cells to ``gene sentences", which is inspired by similarities between natural language system and gene system. Then genes in these sentences are represented as gene embeddings to reduce data sparsity. With these embeddings, we implement a semi-supervised learning model based on recurrent convolutional neural networks (RCNN), which includes a shared network, a supervised network and an unsupervised network. The proposed model is evaluated on macosko2015, a large scale single-cell transcriptomic dataset with ground truth of individual cell types. It is observed that the proposed model is able to achieve encouraging performance by learning on very limited amount of labeled scRNAseq cells  together with a large number of unlabeled scRNAseq cells.
\end{abstract}


%
\IEEEpeerreviewmaketitle

\section{Introduction}
\label{sec1}


Single-cell RNA sequencing (scRNAseq) enables the profiling of the transcriptomes of individual cells, thus characterizing the heterogeneity of biological samples since scRNAseq experiments is able to yield high volumes of data. For example, in a single experiment, the expression profile is up to $10^{5}$ cells, at the level of the single cell~\cite{trapnell2015defining}. It is not possible for traditional bulk RNAseq~\cite{butler2018integrating} to examine biological samples in such high-resolution.  

One common goal of scRNAseq data analytics is to identify the cell type of each individual cell that has been profiled. Although labeling cells with known cell types is a supervised learning task, it is currently achieved by unsupervised methods with manual input~\cite{lieberman2018castle}. To accomplish this, cells are first grouped into different clusters in an unsupervised manner~\cite{ma2019actinn}, and the number of these clusters allows us to approximately determine how many distinct cell types are present. To attempt to interpret the identity of each cluster, marker genes are identified as those that are uniquely highly expressed in a cluster, compared to all other clusters. These canonical markers are then used to assign the cell types for the clusters by cross referencing the markers with lists of previously characterized cell type specific markers. However, this approach has several limitations, including the fact that the clusters may not optimally separate  single cell types, and certain cell types may not have previously characterized markers.  Moreover, these methods are computationally intensive, especially when the number of cells becomes large. 

\begin{figure} []
	\begin{center}
		\includegraphics[width=1.05\linewidth]{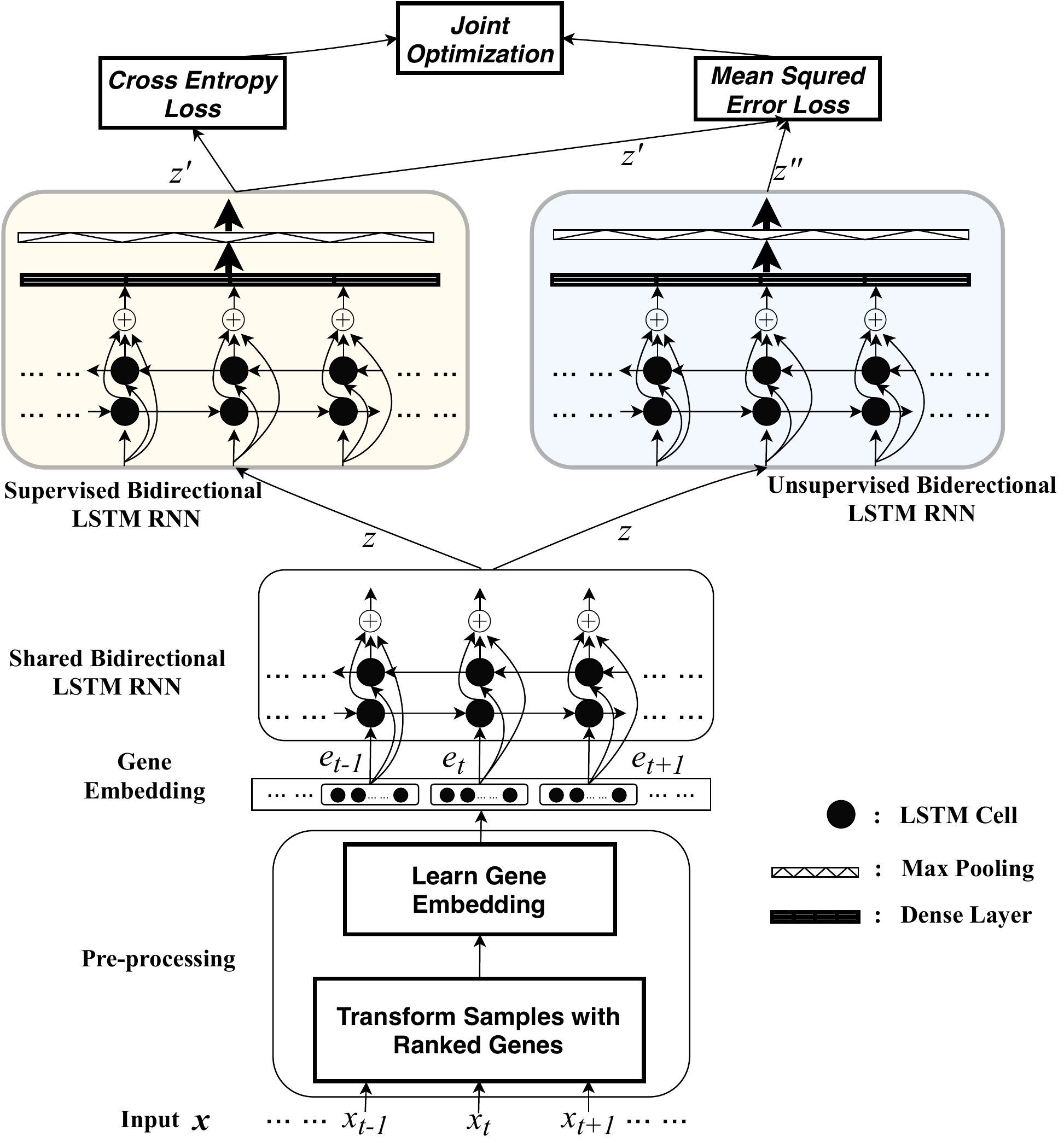}
		\caption{Framework of the proposed semi-supervised learning. Input $x$ is the cell. Cell types are available only for the labeled inputs and the associated cross-entropy loss component is evaluated only for those. $z'$ and $z''$ are outputs from the supervised bidirectional LSTM RNN and the unsupervised bidirectional LSTM RNN, respectively.   We jointly optimize cross entropy loss and mean squared error loss for supervised learning and unsupervised learning with these outputs. $\oplus$ is the concatenation operation. }
		\label{Fig1_architecture}
	\end{center}
\end{figure}

Recently, novel computational methods based on neural networks have been proposed to overcome these limitations~\cite{ lieberman2018castle, ma2019actinn},  since cell type classification based on a large number of genes is much more robust to noise with machine learning models. For example, Ma \textit{et al.} proposed ACTINN (Automated Cell Type Identification using Neural Networks)~\cite{ma2019actinn} with simple neural networks of three neuron layers, which trains on datasets with predefined cell types and predicts cell types for other datasets based on the trained model. It uses all the genes to capture the features for each cell type instead of relying on a limited number of canonical markers. Furthermore, it is much more computationally efficient than traditional approaches. However, it requires a large amount of individual cells with accurate and unbiased type labels to build datasets for training and testing. 

In this paper, \emph{we propose a novel deep semi-supervised learning model when only very limited number of cells are labeled, and a large number of cells are unlabeled}. The proposed framework is shown in Figure~\ref{Fig1_architecture}. It is trained on cells with predefined cell types and then can be used to predict cell types on new datasets. The cells in scRNAseq data are transformed to ``gene sentences" by taking advantage of similarities between natural language system and gene system. Furthermore, to overcome data sparsity, we employ word embedding techniques~\cite{mikolov2013distributed} to represent the genes in these sentences as gene vectors. Then, these vectors are input into the proposed semi-supervised neural networks built on recurrent convolutional neural networks (RCNN)~\cite{lai2015recurrent}. It consists of three components, namely, a shared bidirectional Long Short-Term Memory Recurrent Neural Network (LSTM RNN), a supervised bidirectional LSTM RNN, and an unsupervised bidirectional LSTM RNN. One path is composed of the shared bidirectional LSTM RNN and supervised bidirectional LSTM RNN while the other path consists of the shared bidirectional LSTM RNN and unsupervised bidirectional LSTM RNN. All data (labeled and unlabeled data) will be evaluated to generate the mean squared error loss, while only labeled data will be evaluated to calculate the cross entropy loss.  Experimental results on macosko2015~\cite{macosko2015highly} demonstrate the effectiveness of the proposed model even when training it with very limited amount of labeled cells.

The contributions in this study are as follows.

\begin{itemize}
\item We represent cells in scRNAseq data via embedding techniques to reduce the sparsity of gene expression values.
\item We propose semi-supervised deep learning models with RCNN through jointly training supervised RCNN and unsupervised RCNN. It is shown that the proposed model can learn on unlabeled cells and labeled cells jointly to identify cell types with high performance.
\item The proposed model is validated on a large-scale scRNAseq dataset. Experimental results indicate that the new representations of cells enable cell type identification with promising performance. Moreover, the proposed semi-supervised learning model is able to effectively identify cell types by learning on very limited number of labeled cells and a large amount of unlabeled cells.
\end{itemize}

\section{Problem Formulation}
\label{sec2}

Cell type identification on single-cell transcriptomic data is to classify the individual cells into predefined cell types, which is a supervised learning task from machine learning point of view. Specifically, it is a multi-class classification problem with $N$ cell types in the set $C = \{c_{1}, c_{2}, c_{3}, ..., c_{N}\}$, where $N > 2$. Each cell belongs to one of the $N$ different types. The goal is to construct a function which, given a new individual cell, will correctly predict the cell type where the new individual cell belongs. It is defined by
\begin{equation}
	f(x; \theta) \rightarrow c \; ,
\label{Equ_multi-class}
\end{equation}
where $x$ is an individual cell, $\theta$ denotes the parameters in  $f(\cdot)$, and $c \in C $. For the scRNA-seq data, $x$ is composed of a sequence of gene expression values of the cell. Generally we will have more than $10,000$ gene expression values if we employ scRNAseq techniques to generate data~\cite{lieberman2018castle, ma2019actinn}. These gene expression values will be input as features to build machine learning models to complete cell type identification. Due to high dimensions and data sparsity of the scRNAseq data~\cite{zheng2019emerging}, it is the challenging to solve this problem.

\section{Proposed Methodology}
\label{sec2}

We propose a semi-supervised recurrent convolutional neural network (SSRCNN) to address the challenge of lacking  labeled individual cells for cell type identification from scRNAseq data. The proposed model is based on RCNN~\cite{lai2015recurrent} and the detailed architecture is shown in Figure~\ref{Fig1_architecture}. The first step is to preprocess the scRNAseq data to reduce the data sparsity~\cite{zheng2019emerging, lahnemann2020eleven} by building ``gene sentences" and representing the gene with word embedding techniques~\cite{mikolov2013distributed, goldberg2014word2vec}. Specifically, each cell in the scRNAseq data is composed of thousands of gene expression values. Unfortunately, most of these values are zeros because of the limitation of current single-cell sequencing techniques~\cite{lahnemann2020eleven}, which would reduce the performance of machine learning models significantly~\cite{tran2017missing, zhou2018trajectory}. Therefore, it is important to solve the data sparsity problem for cell type identification.

To overcome the data sparsity, we propose a new technique of ``gene embedding'', to represent the cells based on similarities between gene system and natural language system since gene sequences can be treated as a language when we regards genome as the ``book of life"~\cite{searls2002language}. For example, words can be combined with others to generate new functions ``phrases" while different genes can form pathways to control protein generation~\cite{lesk2017introduction}.  With respect to these similarities, we build gene sentences by selecting $k$ genes and employ word2vec~\cite{mikolov2013efficient} to represent these genes, where word2vec is a powerful technique to overcome data sparsity for natural language processing and understanding~\cite{mikolov2013efficient, yang2013word, schnabel2015evaluation}.  We rank the genes in terms of their expression values and select the top $k$ genes to build the gene sentence. Then the genes in the gene sentence are represented as gene embeddings. For instance, the gene sentence $<$$g_{1}, g_{2}, g_{3}, ..., g_{t}, ..., g_{n}$$>$ will be represented as a sequence of gene embedding $<$$e_{1}, e_{2}, e_{3}, ..., e_{t}, ..., e_{n}$$>$, where $e_{t}$ is the embedding representation of the gene $g_{t}$.  

After the preprocessing procedure, these gene sentences with gene embeddings will be input to the shared bidirectional LSTM RNN to extract common features for cell identification. The forward layer and backward layer generate two directional correlation features, respectively. Next, we combine these two groups of features with the gene embedding and obtain the output $z$  of the shared RNN, where $z$ is a sequence $<$$z_{1}, z_{2}, z_{3}, ..., z_{t}, ..., z_{n}$$>$ and $z_{t}$ is given by
\begin{equation}
	z_{t} =  h_{t}^{f} \oplus e_{t} \oplus h_{t}^{b} \; ,
\label{Equ_output_shared_rnn}
\end{equation}
where 

\begin{equation}
	h_{t}^{f} =  a(w_{h}^{f}h_{t - 1}^{f} + w_{e}^{f}e_{t} + b_{h}^{f}) \; ,
\label{Equ_output_h_f_rnn}
\end{equation}
\begin{equation}
	h_{t}^{b} =  a(w_{h}^{b}h_{t + 1}^{b} + w_{e}^{b}e_{t} + b_{h}^{b}) \; ,
\label{Equ_output_h_b_rnn}
\end{equation}

$z_{t}$ is the output of $g_{t}$ of the gene sentence $<$$g_{1}, g_{2}, g_{3}, ..., g_{t}, ..., g_{n}$$>$. $\oplus$ is the concatenation operation. $a(\cdot)$ is the activation function for hidden layers.  $w_{h}^{f}$ and $w_{e}^{f}$ are forward weights for two layers, namely, forward layer and backward layer. $w_{h}^{b}$ and $w_{e}^{b}$ are backward weights for these two layers, respectively.  $b_{h}^{f}$ and $b_{h}^{b}$ are bias for these two layers. 

The idea to introduce this shared RNN to the proposed model is motivated by deep multi-task learning~\cite{zhang2014facial, ruder2017overview},  since different tasks share a common feature representation based on the original features. In addition, the reason for learning common feature representations instead of directly using the original ones is that the original representation may not have enough expressive power for multiple tasks. With the training data in all tasks, a more powerful representation can be learned for all the tasks and this representation will improve performance.
Therefore, the output $z$ from the shared RNN are evaluated by two bidirectional RNNs, namely, supervised bidirectional LSTM RNN and unsupervised bidirectional LSTM RNN. As shown in Figure~\ref{Fig1_architecture}, the structures of these two RNNs are the same to that of shared RNN. For the supervised RNN, it is to learn the deep features of cells when the sample has the label. The output $z'$  of supervised RNN is the sequence $<$$z'_{1}, z'_{2}, z'_{3}, ..., z'_{t}, ..., z'_{n}$$>$, where $z'_{t}$ is given by 
\begin{equation}
	z'_{t} =  max(tanh(w_{sup}z^{tmp'} + b_{sup})) \; ,
\label{Equ_output_supervised_rnn}
\end{equation}
where 
\begin{equation}
	z^{tmp'} = h_{t'}^{f} \oplus z_{t} \oplus h_{t'}^{b} \; ,
\label{Equ_tmp_supervised_output}
\end{equation}
\begin{equation}
	h_{t'}^{f} =  a(w_{h'}^{f}h_{t' - 1}^{f} + w_{sup}^{f}z_{t} + b_{h'}^{f}) \; ,
\label{Equ_output_supervised_h_rnn}
\end{equation}
\begin{equation}
	h_{t'}^{b} =  a(w_{h'}^{b}h_{t' + 1}^{b} + w_{sup}^{b}z_{t} + b_{h'}^{b}) \; ,
\label{Equ_output_supervised_h_rnn}
\end{equation}

We employ the same activation function $a(\cdot)$ for the hidden layers of the supervised bidirectional RNN. $tanh(\cdot)$ is the activation function for the dense layer. $w_{sup}$ and $b_{sup}$ are the weights and a bias between the max-pooling layer  and the dense layer in the supervised RNN. $w_{h'}^{f}$ and $w_{sup}^{f}$ are forward weights for the forward layer and backward layer in the supervised bidirectional RNN. $w_{h'}^{b}$ and $w_{sup}^{b}$ are backward weights for these two layers, respectively. $b_{h'}^{f}$ and $b_{h'}^{b}$ are bias for these two layers, respectively. 

Moreover, we build the unsupervised bidirectional RNN to generate another representation of the input and the output $z''$  is a vector $<$$z''_{1}, z''_{2}, z''_{3}, ..., z''_{t}, ..., z''_{n}$$>$, where $z''_{t}$ is given by
\begin{equation}
	z''_{t} =  max(tanh(w_{unsup}z^{tmp''} + b_{unsup})) \; ,
\label{Equ_output_unsupervised_rnn}
\end{equation}
where 
\begin{equation}
	z^{tmp''} = h_{t''}^{f} \oplus z_{t} \oplus h_{t''}^{b} \; ,
\label{Equ_tmp_unsupervised_output}
\end{equation}
\begin{equation}
	h_{t''}^{f} =  a(w_{h''}^{f}h_{t'' - 1}^{f} + w_{unsup}^{f}z_{t} + b_{h''}^{f}) \; ,
\label{Equ_output_unsupervised_h_rnn}
\end{equation}
\begin{equation}
	h_{t''}^{b} =  a(w_{h''}^{b}h_{t'' + 1}^{b} + w_{unsup}^{b}z_{t} + b_{h''}^{b}) \; ,
\label{Equ_output_unsupervised_h_rnn}
\end{equation}
$w_{unsup}$ and $b_{unsup}$ are the weights and a bias between the max-pooling layer and the dense layer in the unsupervised RNN. $w_{h''}^{f}$ and $w_{unsup}^{f}$ are forward weights for two layers, namely, forward layer and backward layer in the unsupervised bidirectional RNN. $w_{h''}^{b}$ and $w_{unsup}^{b}$ are backward weights for these three layers, respectively. $b_{h''}^{f}$ and $b_{h''}^{b}$ are bias for these two layers, respectively.

We utilize those two vectors $z'$ and $z''$ to calculate the cross entropy loss (CEL) and mean squared error loss (MSEL) for supervised and unsupervised paths, respectively. They are given by 
\begin{equation}
	l^{CEL} =  -\sum y \times log \phi{(z')} \; ,
\label{Equ_loss1}
\end{equation}
\begin{equation}
	l^{MSEL}  = ||z' - z''||^{2}\; ,
\label{Equ_loss2}
\end{equation}
where $y$ is the label for the input and $\phi(\cdot)$ is the softmax activation function. $l^{CEL}$ is the standard cross entropy loss to account for the loss of labeled inputs.
Because training RNNs with dropout regularization and gradient-based optimization is a stochastic process, the two RNNs will have different link weights after training. In other words, there will be differences between the two prediction vectors $z'$ and $z''$ that are from these two RNNs (supervised RNN and unsupervised RNN). These differences can be treated as an error and thus minimizing its mean square error (MSE) is another objective $l^{MSEL}$, in the proposed model. Furthermore, to combine the supervised loss $l^{CEL}$ and unsupervised loss $l^{MSEL}$, we scale the latter by time-dependent weighting function $w(t)$~\cite{laine2016temporal} that ramps up, starting from zero, along a Gaussian curve. The total loss is defined by 
\begin{equation}
	Loss = l^{CEL} + w(t) \times l^{MSEL}\; ,
\label{Equ_loss}
\end{equation}

At the beginning of training, the total loss and the learning gradients are dominated by the supervised loss component, i.e., the labeled data only. At later stage of training, unlabeled data will contribute more than the labeled data. The detailed learning of the proposed model is shown in  Algorithm \ref{Arg1_learning}, where $f_{r}(\cdot)$ is to represent cells as gene sentences, $f_{e}(\cdot)$ is to learn gene embeddings on the gene sentences, and $f_{\theta_{shared}}(\cdot)$ is to learn the common features from the gene embeddings.  Parameters of the shared neural network $\theta_{shared}$  include $w_{h}^{f}$, $w_{b}^{f}$, $w_{e}^{f}$, $w_{e}^{b}$, $b_{h}^{f}$, and $b_{h}^{b}$. 

\begin{algorithm}[h!]
	\caption{Learning of SSRCNN}
	\begin{algorithmic}[1]
		\Require{training sample  $x_{i}$, the set of training samples  $S$, labeled samples $y_{i}$ for $x_{i}$ ($i \in S$) } 
		\For{$t$ in~[1, num epochs] }
 			 \For{each minibatch $B$}
			 	\State{$x'_{i \in B} \gets f_{r}(x_{i \in B})$} $\triangleright$ preprocessing
			 	\State{$x''_{i \in B} \gets f_{e}(x'_{i \in B})$} $\triangleright$  gene embedding
				\State{$z_{i \in B} \gets f_{\theta_{shared}}(x''_{i \in B})$} $\triangleright$ common  feature extraction
      				\State{$z'_{i \in B} \gets f_{\theta_{sup}}({{z_{i \in B}}})$} $\triangleright$ supervised  representation
				\State{$z''_{i \in B} \gets f_{\theta_{unsup}}({{z_{i \in B}}})$} $\triangleright$ unsupervised  representation
				\State{$l_{i \in B}^{CEL} \gets -\frac{1}{|B|} \sum_{i \in B \cap S}{log \phi{(z'_{i})}[y_{i}]}$}  $\triangleright$ supervised loss component
				\State{$l_{i \in B}^{MSEL} \gets \frac{1}{C|B|} \sum_{i \in B}{||z'_{i} - z''_{i}||^{2}}$} $\triangleright$ unsupervised loss component
				\State{$Loss \gets l_{i \in B}^{CEL} + w(t) \times l_{i \in B}^{MSEL}$} $\triangleright$ total loss 
				\State{update $\theta_{shared}$,  $\theta_{sup}$, $\theta_{unsup}$ using, e.g., ADAM} 
 			 \EndFor
		\EndFor
	\Return{$\theta_{shared}$,  $\theta_{sup}$, $\theta_{unsup}$}
	\end{algorithmic}
	\label{Arg1_learning}
\end{algorithm}

After extracting common features from gene samples, we use $f_{\theta_{sup}}(\cdot)$ and $f_{\theta_{unsup}}(\cdot)$ to obtain higher level representations to complete cell type identification and enhance the cell representations through optimizing supervised loss and unsupervised loss jointly. Parameters of the supervised RNN $\theta_{sup}$  include $w_{h'}^{f}$, $w_{h'}^{b}$, $w_{sup}^{f}$, $w_{sup}^{b}$, $b_{h'}^{f}$, $b_{h'}^{b}$, $w^{sup}$, and $b^{sup}$. Parameters of the unsupervised RNN $\theta_{unsup}$  consist of $w_{h''}^{f}$, $w_{h''}^{b}$, $w_{unsup}^{f}$, $w_{unsup}^{b}$, $b_{h''}^{f}$, $b_{h''}^{b}$, $w^{unsup}$, and $b^{unsup}$. 

The proposed model combines the advantages of deep multi-task learning~\cite{zhang2014facial} and $\Pi$ model \cite{laine2016temporal}. However,  there exist significant differences. Compared to deep multi-task learning, the subtasks in the proposed model have two categories of learning, namely, supervised learning and unsupervised learning while there is only supervised learning in the deep multi-task learning. On the other hand, instead of using one path neural networks, we apply two independent RNNs to generate supervised and unsupervised outputs. Furthermore, the proposed model is more flexible as the two independent RNNs can be tuned in terms of specific goals.


\section{Experiment}
\label{sec4}

\subsection{Dataset}

\begin{figure} [h!]
	\includegraphics[width=\linewidth]{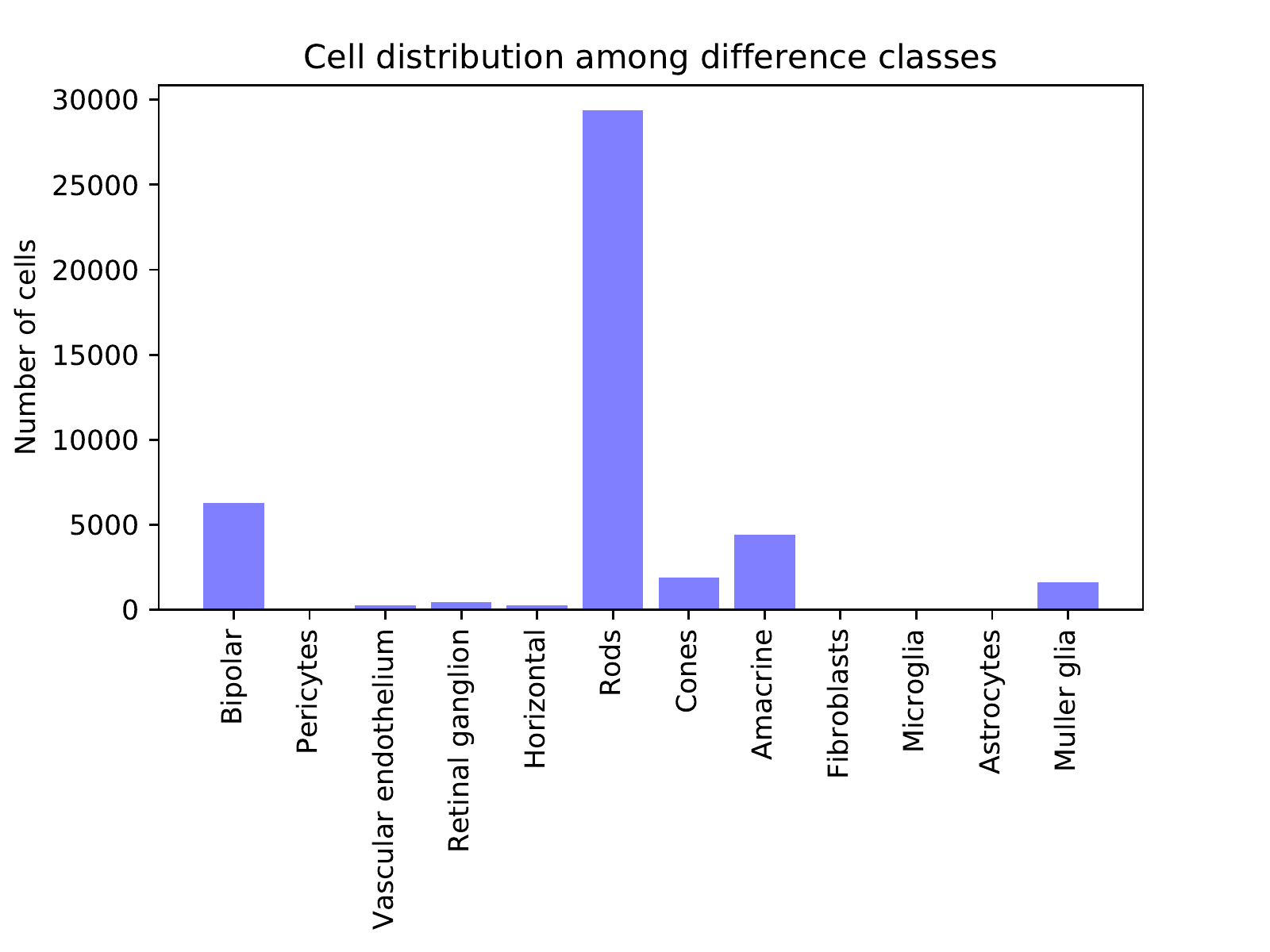}
	\caption{Cell distribution in different types. }
	\label{Fig2_cell}
\end{figure}

We evaluate our proposed method using macosko2015~\cite{macosko2015highly}, a retina scRNAseq dataset. It includes 44,825 mouse retinal cells with 39 transcriptionally distinct cell populations\footnote{https://github.com/olgabot/macosko2015}. The dataset with 24,760 genes contains 12 cell types, namely, rods, cones, muller glia, astrocytes,  fibroblasts, vascular endothelium, pericytes, microglia, retinal ganglion, bipolar, horizontal, and amacrine. The cell type distribution is shown in Figure~\ref{Fig2_cell}.  It can be observed that the cell distribution is imbalanced across different cell types. Therefore, machine learning models built on this data will have bias to majority classes. In other words, the models will tend to obtain high  performance for identification of majority cell types, but low performance for identification of minority cell types. It will be a challenge to implement cell type classification with high performance for all cell types.

\subsection{Experimental settings}
In this experiment, our proposed model is employed to implement  cell type identification. The key hyper parameters of the proposed model are: 
Embedding size: 256
Minibatch size: 128,
Number of epoch: 300,
Optimizer: Adam optimizer,
Learning rate: 0.001,
Learning rate decay: 0.9.
They are determined by trial and error. For the data preprocessing, we select top 50 genes based on the gene expression values to build the gene sentence for each cell. Moreover, the details of the model architecture is illustrated in Table~\ref{Table1_NetworkArchitecture}. Specifically, the output of the proposed model contains two parts: cell type $\phi{(z')}$  and a new representation $z''$.

\begin{table}[h!]
	\caption{The proposed network architecture.}
	\begin{center}
	\begin{tabular}{ll}
		\hline
		\textbf{Name} & \textbf{Description}  \\ \hline
		Input &  Gene Sentence \\ \hline
		Gene Embedding &  Mikolov model~\cite{mikolov2013efficient, mikolov2013exploiting}\\ \hline
		Shared RNN &  256 LSTM cells for each hidden layer, \\
		&  one forward hidden layer, \\& one backward hidden layer\\ \hline
		Supervised RNN &  256 LSTM cells for each hidden layer, \\
		&  one forward hidden layer,  \\ &  one backward hidden layer,\\
		&  one dense layer with 256 neurons,  \\ & one $2\times2$ max-pooling layer\\ \hline
		Unsupervised RNN &  256 LSTM cells for each hidden layer, \\
		&  one forward hidden layer, \\ & one backward hidden layer,\\
		&  one dense layer with 256 neurons,  \\ & one $2\times2$ max-pooling layer\\ \hline
		Output & cell type $\phi{(z')}$  and a new representation $z''$\\ \hline
	\end{tabular}
	\end{center}
	\label{Table1_NetworkArchitecture}
\end{table}

\subsection{Evaluation metric}
We applied different evaluation metrics to evaluate the performance of our proposed model, which includes accuracy, macro-average Precision (MacroP), macro-average Recall (MacroR), and macro-average Fscore (MacroF)~\cite{van2013macro}. Accuracy is calculated by dividing the number of cells identified correctly over the total number of testing cells. 

\begin{equation}
	Accuracy = \frac{N_{correct}}{N_{total}}.
\end{equation}

Macro-average~\cite{yang2001study} is to calculate the metrics such as Precision, Recall and F-scores independently for each cell type and then utilize the average of these metrics.  It is to evaluate the whole performance of classifying cell types.  
\begin{equation}
	MacroF = \frac{1}{C} \sum_{c=1}^{C} Fscore_c.
\end{equation}
\begin{equation}
	MacroP= \frac{1}{C} \sum_{c=1}^{C} Precision_c.
\end{equation}
\begin{equation}
	MacroR= \frac{1}{C} \sum_{c=1}^{C} Recall_c.
\end{equation}
where $C$ denotes the total number of cell types and $Fscore_c$, $Precision_c$, $Recall_c$ are $Fscore$, $Precision$, $Recall$ values in the $c^{th}$ cell type which are defined by

\begin{equation}
	Fscore = \frac{2 \times Precision \times Recall}{Precision + Recall}.
\end{equation}

where $Precision$ indicates precision measurement that defines the capability of a model to represent only correct cell types and $Recall$ computes the aptness to refer all corresponding correct cell types:

\begin{equation}
	Precision = \frac{TP}{TP+FP}.
\end{equation}

\begin{equation}
	Recall = \frac{TP}{TP+FN}.
\end{equation}
whereas ${TP}$ (True Positive) counts total number of cells matched with the cells in the types. ${FP}$ (False Positive) measures the number of recognized type does not match  the annotated cells. ${FN}$ (False Negative) counts the number of cells that does  not match  the predicted cells. 
The main goal for learning from imbalanced datasets such as macosko2015~\cite{macosko2015highly} is to improve the recall without hurting the precision. However, recall and precision goals are often conflicting, since when increasing the true positive (TP) for the minority class (True), the number of false positives (FP) can also be increased; this will reduce the precision~\cite{chawla2009data}. 

In addition, we employ three deep supervised learning models as baselines including 1) Word-level CNN (Word CNN) \cite{kim2014convolutional},  2) Attention-Based Bidirectional RNN (Att RNN)~\cite{zhou2016attention},  and 3) Recurrent CNN (RCNN)~\cite{lai2015recurrent}, where these models perform well on similar problems such as text classification. For example, Word CNN performs well on sentence classification, which is more suitable to process sequencing data as the length of the content of the data is short like that of the gene sentence. In addition, we build 4) word-level bidirectional RNN (Word RNN) to compare the implemented model, where Word RNN contains one embedding layer and one bidirectional RNN layer, and concatenate all the outputs from the RNN layer to feed to the final layer that is a fully-connected layer. Moreover, we employ 6 traditional machine learning models as the baselines, namely, Naive Bayes, Decision Tree, Random Forest, Adaboost, Neural Networks (NN), and Support Vector Machine (SVM). Thus, there are total 10 baseline models. \emph{Note that baseline models are built on all labeled cells from the original training datasets}.

\begin{table*}[h!]
	\caption{\label{tab1_ml} Comparing performance between traditional machine learning (ML) and deep learning (DL).}  
        \begin{center}
                \begin{tabular}{|l|llllll|}
                 \hline \multirow{7}{*}{\textbf{Original Gene Expression}} & \multicolumn{1}{l}{\textbf{Machine Learning (ML)} } & \multicolumn{1}{l}{\textbf{Accuracy}} & \multicolumn{1}{l}{\textbf{MacroP}} & \multicolumn{1}{l}{\textbf{MacroR}} & \multicolumn{1}{l}{\textbf{MacroF}} & \multicolumn{1}{c|}{\textbf{Training Time (s)}} \\\cline{2-7}
                    	  & \multicolumn{1}{l}{Naive Bayes}				&  \multicolumn{1}{c}{35.06\%}		& \multicolumn{1}{c}{36.96\%}		& \multicolumn{1}{c}{30.40\%}		&  \multicolumn{1}{c}{35.48\%}  	&  \multicolumn{1}{c|}{11} \\                      
                           & \multicolumn{1}{l}{Random Forest}			&  \multicolumn{1}{c}{85.09\%}		& \multicolumn{1}{c}{55.44\%}		& \multicolumn{1}{c}{27.45\%}		&  \multicolumn{1}{c}{31.03\%} 		&  \multicolumn{1}{c|}{22}  \\                          
                           & \multicolumn{1}{l}{Neural Networks}			&  \multicolumn{1}{c}{86.72\%}		& \multicolumn{1}{c}{19.47\%}		& \multicolumn{1}{c}{23.77\%}		&  \multicolumn{1}{c}{21.23\%} 		&  \multicolumn{1}{c|}{187}  \\
                           & \multicolumn{1}{l}{Decision Tree}				&  \multicolumn{1}{c}{93.78\%} 		& \multicolumn{1}{c}{86.60\%}		& \multicolumn{1}{c}{80.34\%}		&  \multicolumn{1}{c}{82.69\%} 		&  \multicolumn{1}{c|}{1,172}  \\
                           & \multicolumn{1}{l}{Adaboost	}				&  \multicolumn{1}{c}{74.07\%}		& \multicolumn{1}{l}{30.38\%}		& \multicolumn{1}{c}{26.88\%}		&  \multicolumn{1}{c}{25.67\%} 		&  \multicolumn{1}{c|}{1,767}  \\
                           & \multicolumn{1}{l}{\textbf{Support Vector Machine}}	&  \multicolumn{1}{c}{\textbf{97.28\%}} 	& \multicolumn{1}{c}{ \textbf{98.24\%}	}	& \multicolumn{1}{c}{\textbf{93.32\%}}	& \multicolumn{1}{c}{\textbf{95.50\%}} &  \multicolumn{1}{c|}{\textbf{5,554}} \\

                   \hline
                    \hline \multirow{5}{*}{\textbf{Gene Embedding}} & \multicolumn{1}{l}{\textbf{Deep Learning (DL)} } & \multicolumn{1}{l}{\textbf{Accuracy}} & \multicolumn{1}{l}{\textbf{MacroP}} & \multicolumn{1}{l}{\textbf{MacroR}} & \multicolumn{1}{l}{\textbf{MacroF}} & \multicolumn{1}{l|}{\textbf{Training Time (s)}} \\\cline{2-7}
                    	   & \multicolumn{1}{l}{Word CNN~\cite{kim2014convolutional}}		&  \multicolumn{1}{c}{96.30\%}		& \multicolumn{1}{c}{90.79\%}		& \multicolumn{1}{c}{77.22\%}		&  \multicolumn{1}{c}{81.90\%}  	&  \multicolumn{1}{c|}{295}  \\
                            & \multicolumn{1}{l}{Word RNN}							&  \multicolumn{1}{c}{96.11\%}  	& \multicolumn{1}{c}{86.69\%}		& \multicolumn{1}{c}{82.82\%}		&  \multicolumn{1}{c}{84.17\%} 		&  \multicolumn{1}{c|}{8,368}  \\
                            & \multicolumn{1}{l}{Attenion RNN~\cite{zhou2016attention}}		&  \multicolumn{1}{c}{95.79\%}		& \multicolumn{1}{c}{88.18\%}		& \multicolumn{1}{c}{84.85\%}		&  \multicolumn{1}{c}{85.85\%} 		&  \multicolumn{1}{c|}{4,661}  \\
                            & \multicolumn{1}{l}{\textbf{RCNN~\cite{lai2015recurrent}}}	&  \multicolumn{1}{c}{\textbf{96.56\%}}	& \multicolumn{1}{c}{\textbf{96.55\%}}		& \multicolumn{1}{c}{\textbf{92.70\%}}		&  \multicolumn{1}{c}{\textbf{94.45\%}} 	&  \multicolumn{1}{c|}{\textbf{2,383}}  \\
                     \hline                 
                \end{tabular}
       \end{center}
         \label{tab1_gs}
\end{table*}

\subsection{Experimental results}

We evaluated the proposed model from two perspectives. One is to verify if the data preprocessing of the cell is able to be employed to identify cell types effectively. The other is to validate performance of the proposed model on cell type identification with limited amount of labeled cells.

\subsubsection{Data preprocessing}
Table~\ref{tab1_gs} presents the comparison of identification performance between traditional machine learning (ML) models and deep learning (DL)  models, where the ML models are built on the original gene values without data preprocessing while the DL models are built on preprocessed data that includes gene sentences with gene embeddings.

We can observe that most of ML models perform not well on the cell identification regarding the data sparsity. For example, Naive Bayes's accuracy and MacroF are not high since it is sensitive to data sparsity and cell imbalance. Other four ML including Decision Tree, Random Forest, Adaboost and NN identify cell type with high accuracy but low MacroF since they cannot overcome the challenge of cell imbalance even if data sparsity will not affect their performance significantly. Only SVM can perform well on accuracy and MacroF. However, it will cost almost one and a half hours to obtain a converged model with respect to training on such a big scRNAseq data. 

On the contrary, different DL models built on preprocessed cell data can identify cell types with promising and consistent performance. For instance, compared to ML models, all DL models are able to gain high accuracy above 95\%, which means they are not struggling to the data sparsity. Moreover, considering MacroF values, DL models can obtain encouraging performance since these models can overcome cell imbalance to some extent. Specifically, the performance difference between RCNN and SVM is not significant regarding accuracy and MacroF. Moreover, compared to SVM, building RCNN only uses about a half of hour to become converged. Based on the observations, we believe that the preprocessing is an effective step to prepare the data for deep learning based cell type identification.

\subsubsection{Cell type identification}

In this session, we will examine if the proposed model is able to effectively identify the cell types by training on very limited amount of annotated cells. Table~\ref{tab2_gs} presents the comparison of identification performance between SVM, RCNN, and the proposed model, where the proposed model is built based on RCNN with different ratios of training labeled cells. Firstly, we observe that the performance of proposed model is enhanced through increasing the ratios of annotated cells. In other words, the proposed model is able to obtain stronger identification ability when learning on more labeled data. It's because the unsupervised path is able to enhance the data representation for improving cell identification that is implemented with supervised path. 

\begin{table}[h!]
	\caption{\label{tab1_ml} Comparing performance between SVM, RCNN, and Our model (Semi-supervised recurrent convolutional neural networks, SSRCNN). }
       
        \begin{center}
                \begin{tabular}{|l|cccc|}
                \hline \textbf{ML} & \textbf{Accuracy} & \textbf{MacroP} & \textbf{MacroR} & \textbf{MacroF}\\ \hline
                           SVM		&  97.28\%	& 98.24\%		& 93.32\%		& 95.50\% \\  
                \hline       
                 \hline \textbf{DL} & \textbf{Accuracy} & \textbf{MacroP} & \textbf{MacroR} & \textbf{MacroF}\\ \hline
                            RCNN~\cite{lai2015recurrent}			&  96.56\%	& 96.55\%		& 92.70\%		&  94.45\% \\
                   \hline
                    \hline \textbf{Our model} & \textbf{Accuracy} & \textbf{MacroP} & \textbf{MacroR} & \textbf{MacroF}\\ \hline
                    	   SSRCNN (1\%)	&  95.47\%	& 91.73\%		& 93.90\%		&  92.64\%  \\
	   		   SSRCNN (3\%)	&  95.76\%	& 92.62\%		& 94.21\%		&  93.28\%  \\
			   SSRCNN (5\%)	&  95.76\%	& 93.12\%		& 93.39\%		&  93.18\%  \\
	   		   SSRCNN (10\%)	&  95.70\%	& 94.92\%		& 93.18\%		&  93.87\%  \\
                    	   SSRCNN (30\%)	&  96.44\%	& 96.53\%		& 92.66\%		&  94.46\%  \\
                     \hline
                      
                \end{tabular}
       \end{center}
         \label{tab2_gs}
\end{table}

Compared to SVM and RCNN, the proposed model can identify the cell types even with extremely small amount of annotated cells. For example, we can obtain encouraging performance with 1\% annotated cells. Furthermore, the proposed model is robust since we can gain similar performance with different ratios of annotated cells. For instance, the differences of accuracy and MacroF  between the case of 1\%, 5\%, and 30\% are about 1\%. Specifically, the MacroP is improved significantly when increasing the ratios of labeled cells for training while the MacroR is stable. The reason for this observation is that enhancing representation with unsupervised learning in the proposed model seems to be more useful to identify cell type precisely. 

In addition to examining the performance comparisons between the proposed models and baselines, we have to figure out whether the proposed model is sensitive the hyper-parameters.  There are various hyper-parameters involved in the learning procedure of the proposed model. Here, we choose batch size to check since different batch sizes will involve different numbers of labeled cells for building the proposed model when using the same ratio of labeled cells. Table~\ref{tab_batch_size} shows the comparison results for two different batch sizes. We observe that there is no significant differences of the performance. It means that the proposed model is not sensitive to the batch size since the supervised and unsupervised RNN in the proposed model could collaborate with each other to overcome the effects from the difference of batch size.
\begin{table}[h!]
	\caption{\label{tab_batch_size} Comparing performance with different batch sizes on different labeled ratios.} 
        \begin{center}
                \begin{tabular}{|c|cccc|}
                    \hline  &  & 1\% Labeled Data &  & \\ 
                    \hline \textbf{Batch size} & \textbf{Accuracy} & \textbf{MacroP} & \textbf{MacroR} & \textbf{MacroF}\\ \hline
                    	   128			&  95.47\%	& 91.73\%		& 93.90\%		&  92.64\% \\
                            256			&  95.11\%  	& 89.99\%		& 94.40\%		&  91.88\% \\
                     \hline
                    \hline  &  & 3\% Labeled Data &  & \\
                    \hline \textbf{Batch size} & \textbf{Accuracy} & \textbf{MacroP} & \textbf{MacroR} & \textbf{MacroF}\\ \hline
                    	   128			&  95.76\%	& 92.62\%		& 94.21\%		&  93.28\% \\
                            256			&  95.44\%  	& 91.76\%		& 94.21\%		&  92.79\% \\
                      \hline
                     \hline  &  & 5\% Labeled Data &  & \\ 
                      \hline \textbf{Batch size} & \textbf{Accuracy} & \textbf{MacroP} & \textbf{MacroR} & \textbf{MacroF}\\ \hline
                    	   128			&  95.76\%	& 93.12\%		& 93.39\%		&  93.18\% \\
                            256			&  95.49\%  	& 91.34\%		& 93.74\%		&  92.31\% \\
                   \hline
                    \hline  &  & 10\% Labeled Data &  & \\
                    \hline \textbf{Batch size} & \textbf{Accuracy} & \textbf{MacroP} & \textbf{MacroR} & \textbf{MacroF}\\ \hline
                    	   128			&  95.70\%	& 94.92\%		& 93.18\%		&  93.87\% \\
                            256			&  95.93\%  	& 95.13\%		& 93.11\%		&  94.00\% \\
                      \hline
                     \hline  &  & 30\% Labeled Data &  & \\ 
                      \hline \textbf{Batch size} & \textbf{Accuracy} & \textbf{MacroP} & \textbf{MacroR} & \textbf{MacroF}\\ \hline
                    	   128			&  96.44\%	& 96.53\%		& 92.66\%		&  94.46\% \\
                            256			&  96.45\%  	& 96.58\%		& 92.02\%		&  94.10\% \\
                   \hline
                \end{tabular}
       \end{center}
         \label{tab_batch_size}
\end{table}

To further investigate the detailed performance, we show the performance with confusion matrix. Figure~\ref{Fig3_cell} presents the confusion matrix on performance generated with different ratios of annotated cells when using batch size 128 to build the proposed model. It is observed that for different cell types, the accuracy is increased when involving more labeled cells to build the model.  Specifically, when we use different ratios of labeled cells to build the model, the error distributions are not changed significantly. For instance, for the cell type $c_{2}$, the majority errors are from incorrectly classifying the cells into the cell type $c_{7}$.

\newcommand{\photo}[1]
{
    \includegraphics[width=7cm]{#1}
}
Furthermore, considering the unbalanced feature of cell distribution (See Figure \ref{Fig2_cell}), the results in Figure~\ref{Fig3_cell} presents the model bias for the minority cell types. It means that the model will obtain higher performance for the majority type, but lower performance for the minority types. For the cell type $c_{12}$, compared to the case of 1\% labeled cells,  the accuracy is decreased because of the model bias when using 10\% labeled cells for training. 

\begin{figure*}[h!]
 \begin{center}
\begin{tabular}{cc}
	\photo{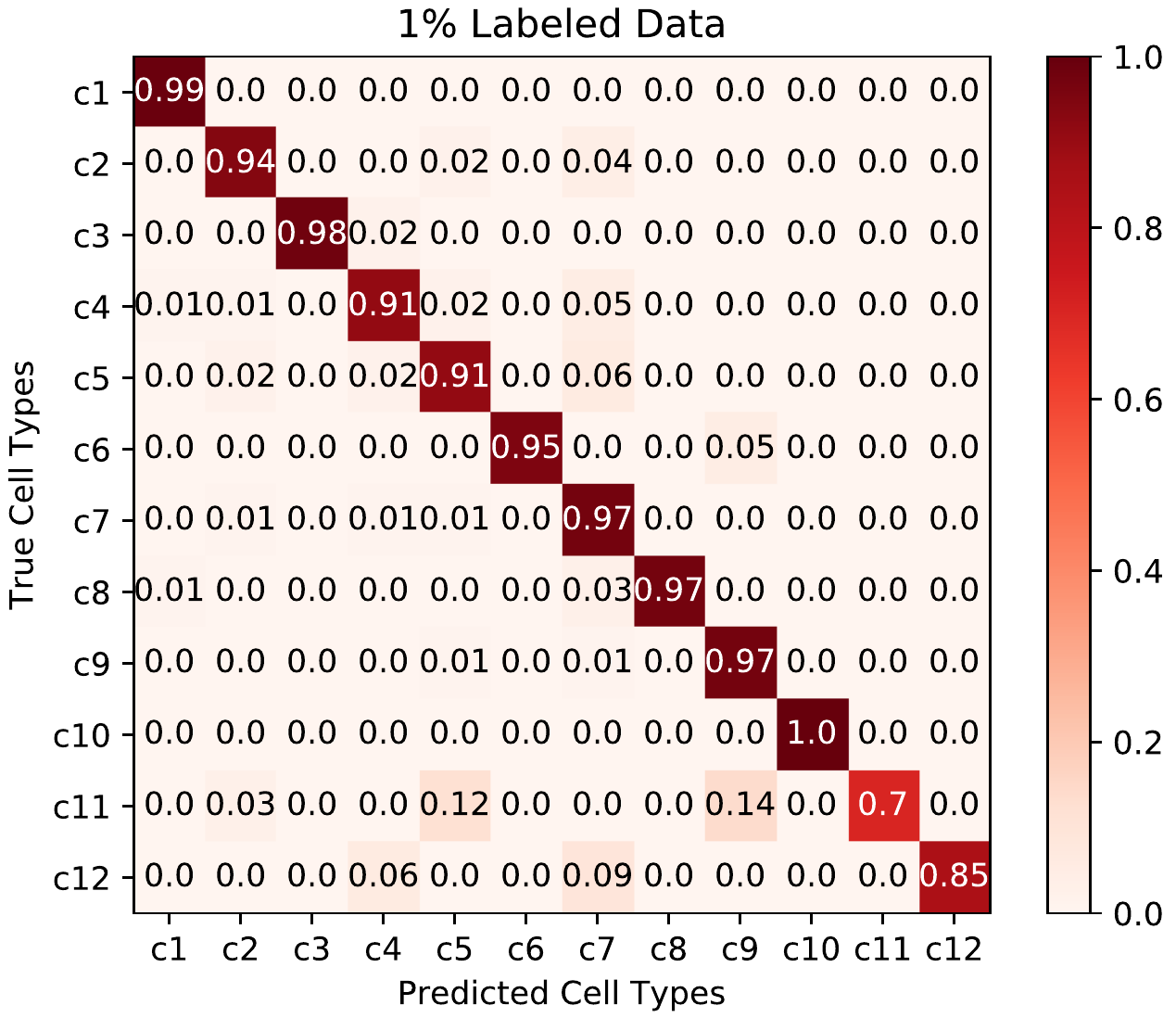} &  \photo{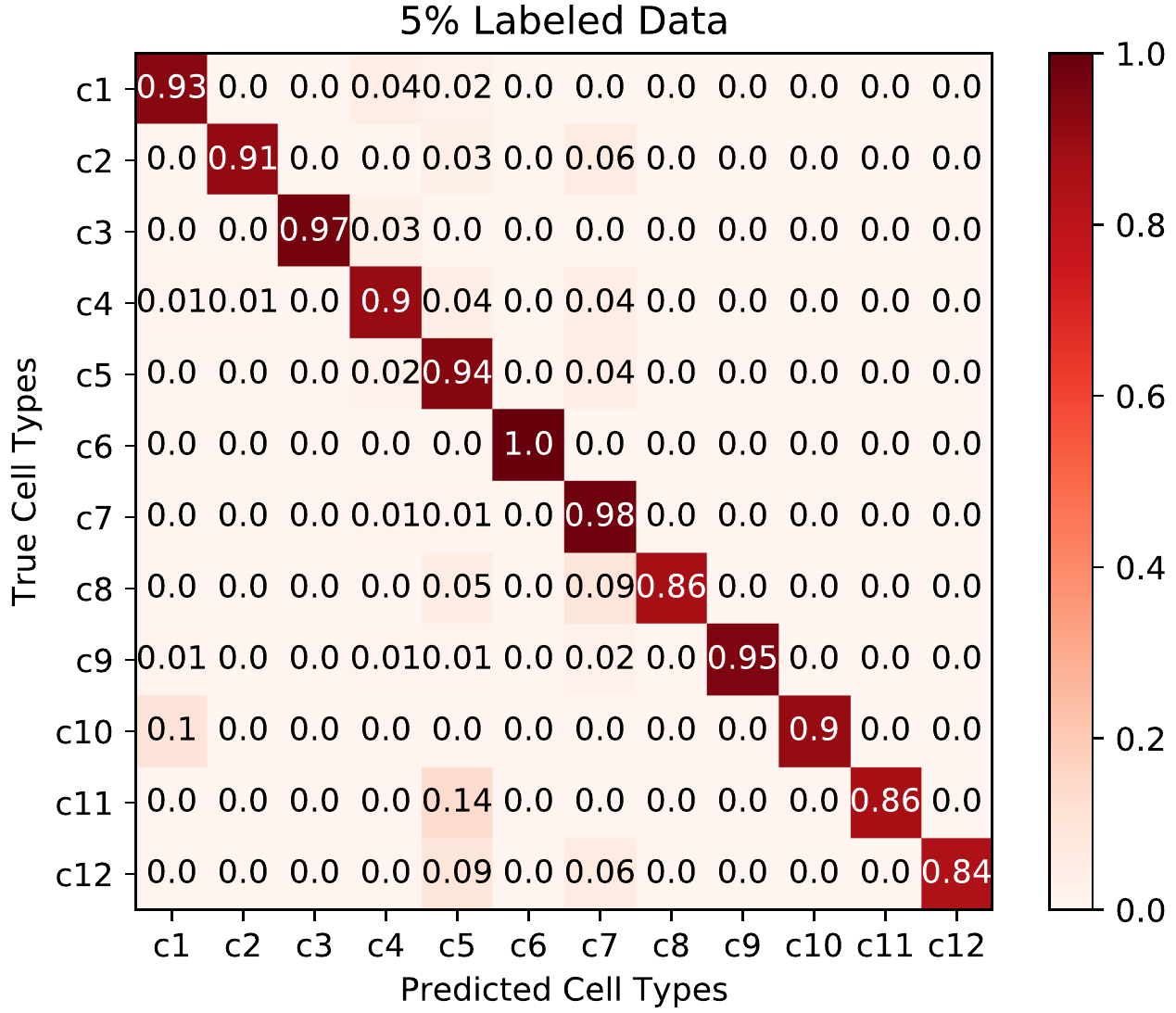} \\
	\photo{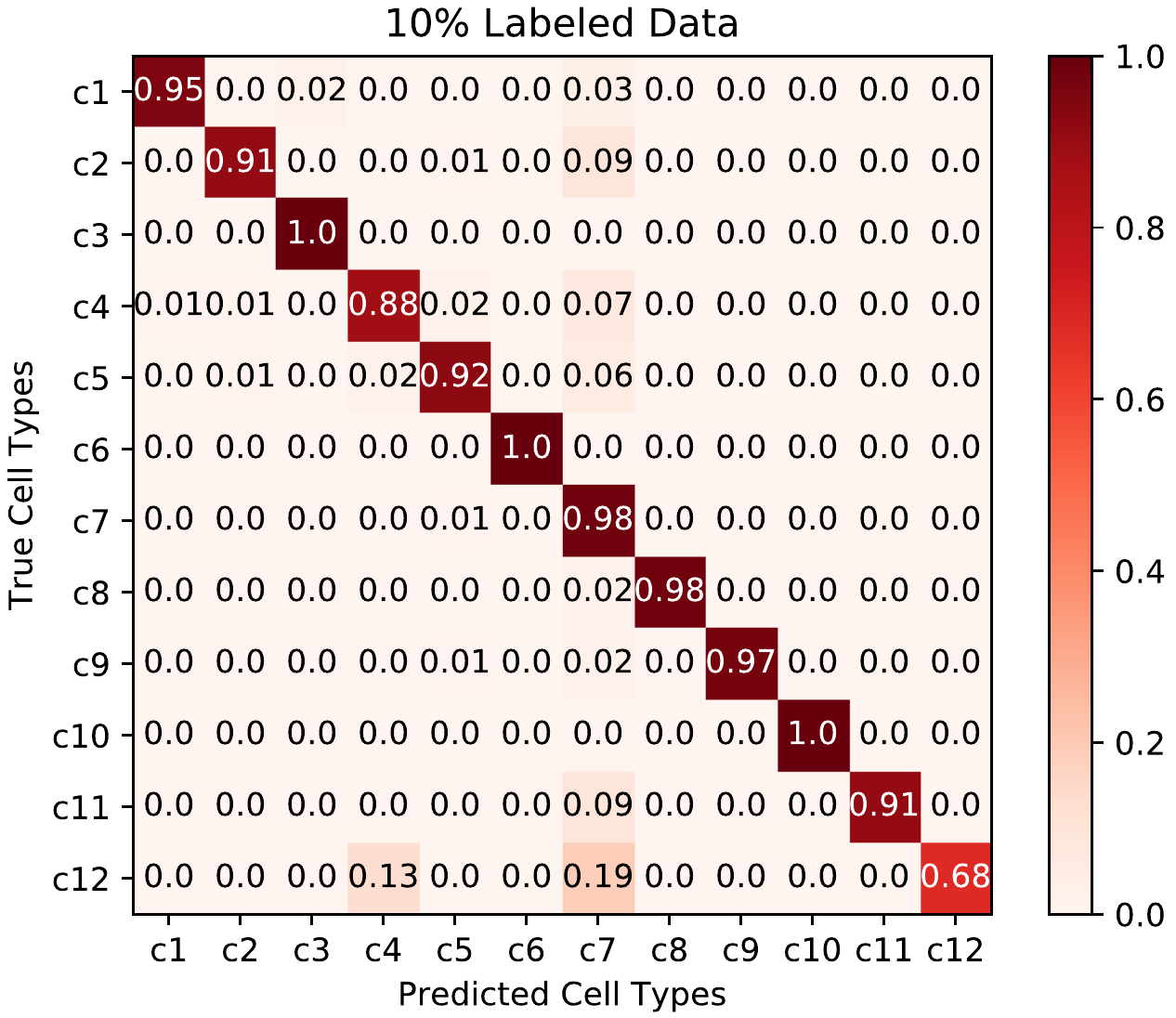} &  \photo{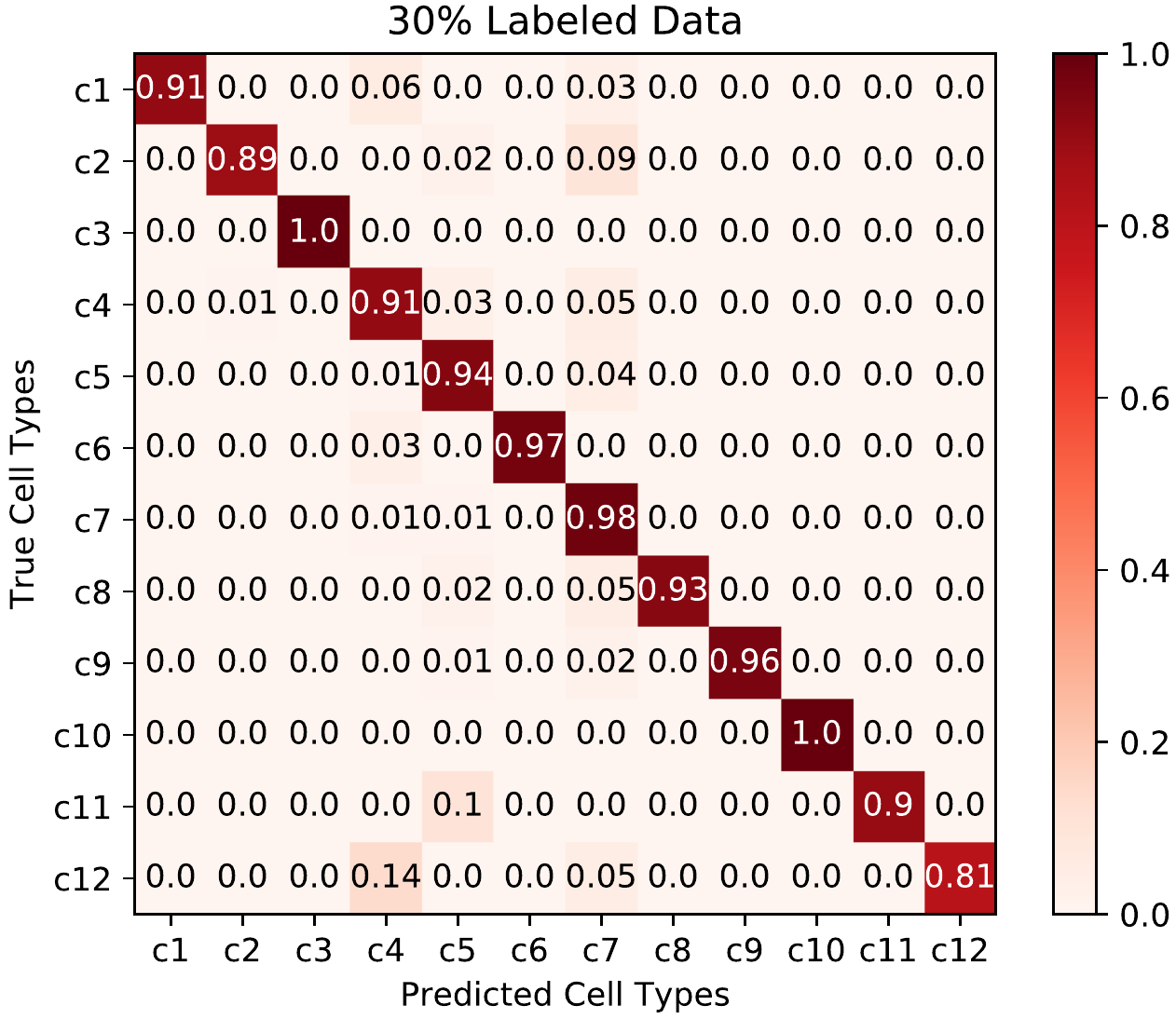} \\	
\end{tabular}
 \end{center}
 \caption{Confusion matrix on different cell types generated with batch size 128. There are 12 cell types including $c_{1}$ (Bipolar), $c_{2}$ (Pericytes), $c_{3}$ (Vascular endothelium), $c_{4}$ (Retinal ganglion), $c_{5}$ (Horizontal), $c_{6}$ (Rods), $c_{7}$ (Cones), $c_{8}$ (Amacrine), $c_{9}$ (Fibroblasts), $c_{10}$ (Microglia), $c_{11}$ (Astrocytes), $c_{12}$ (Muller glia) }
 \label{Fig3_cell}
\end{figure*}

On the other hand, although the overall prediction accuracy (See Table \ref{tab2_gs}) is increased when increasing the ratios of labeled cells, it is not always true that the accuracy for each cell type will be enhanced. This can be observed in Fig~\ref{Fig3_cell}. Take the cell type $c_{12}$ as an example, the prediction accuracy is not always increased when increasing the ratios of labeled cells. On the contrary,  for the cell type $c_{11}$, the accuracy is improved whenever more labeled cells are involved for building the identification model. 

\begin{figure*}[h!]
 \begin{center}
\begin{tabular}{cc}
	\photo{Figure/result/128/Figure_3-1.pdf} &  \photo{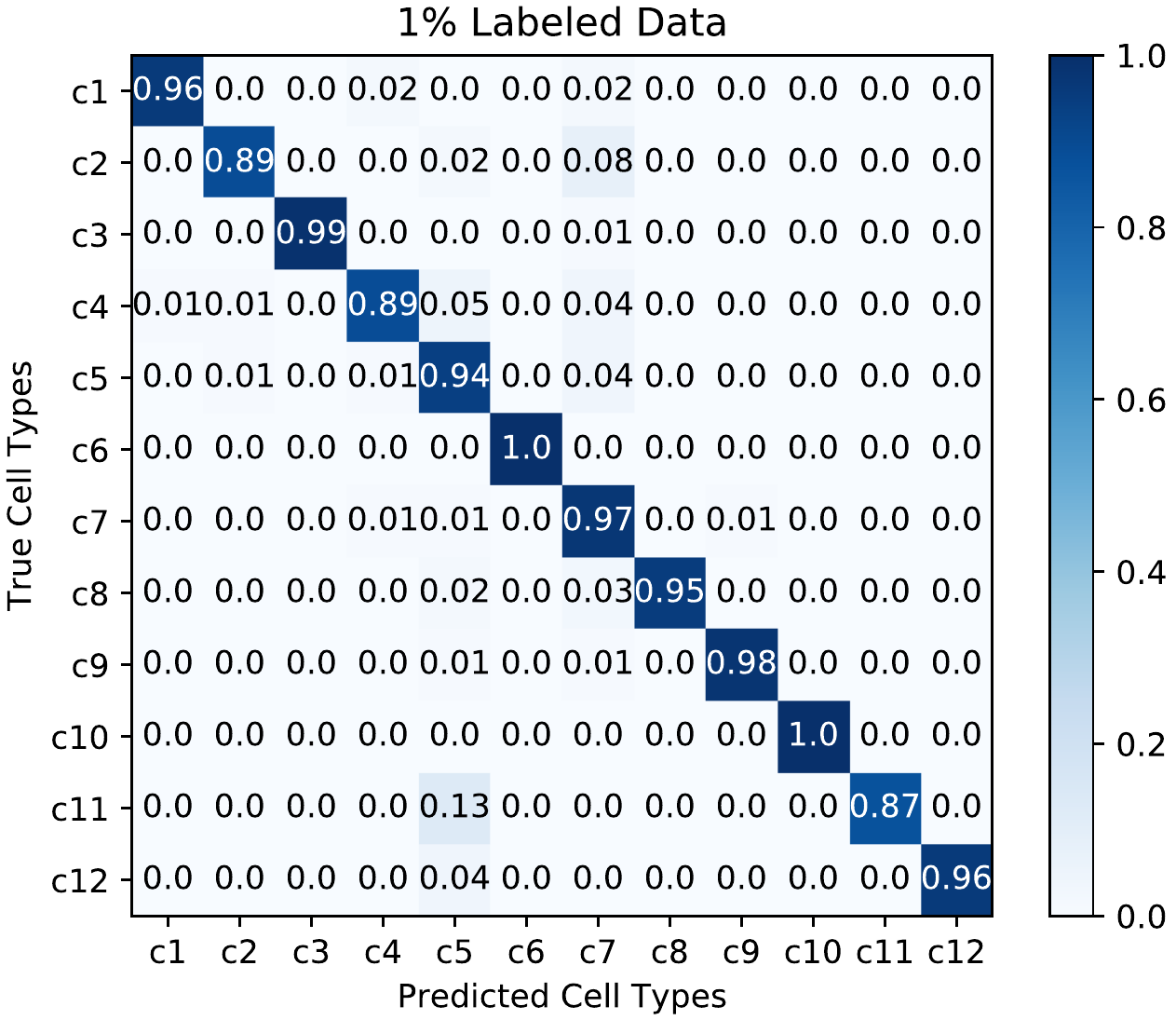} \\
	\photo{Figure/result/128/Figure_3-3.pdf} &  \photo{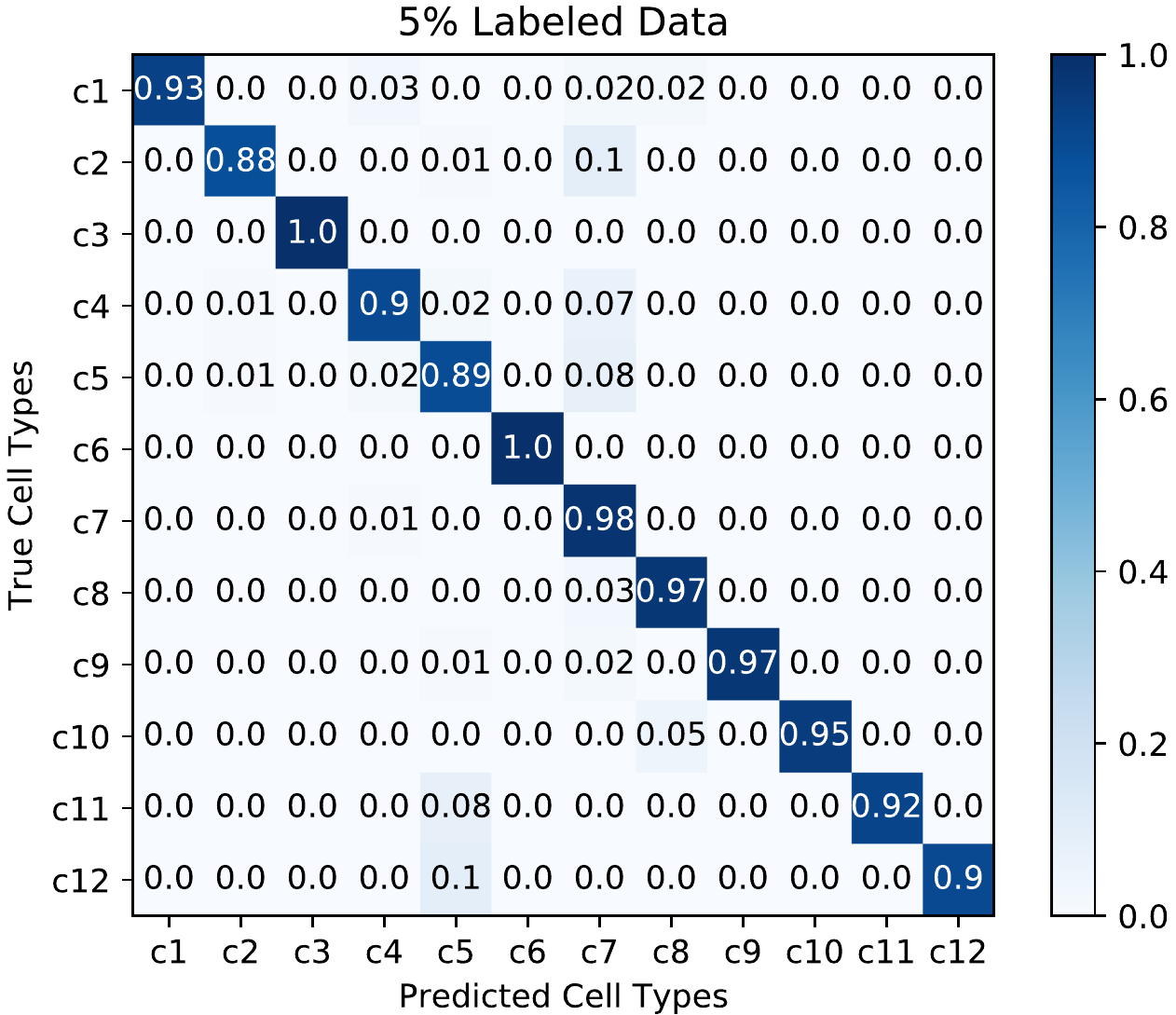} \\	
	\photo{Figure/result/128/Figure_3-5.pdf} &  \photo{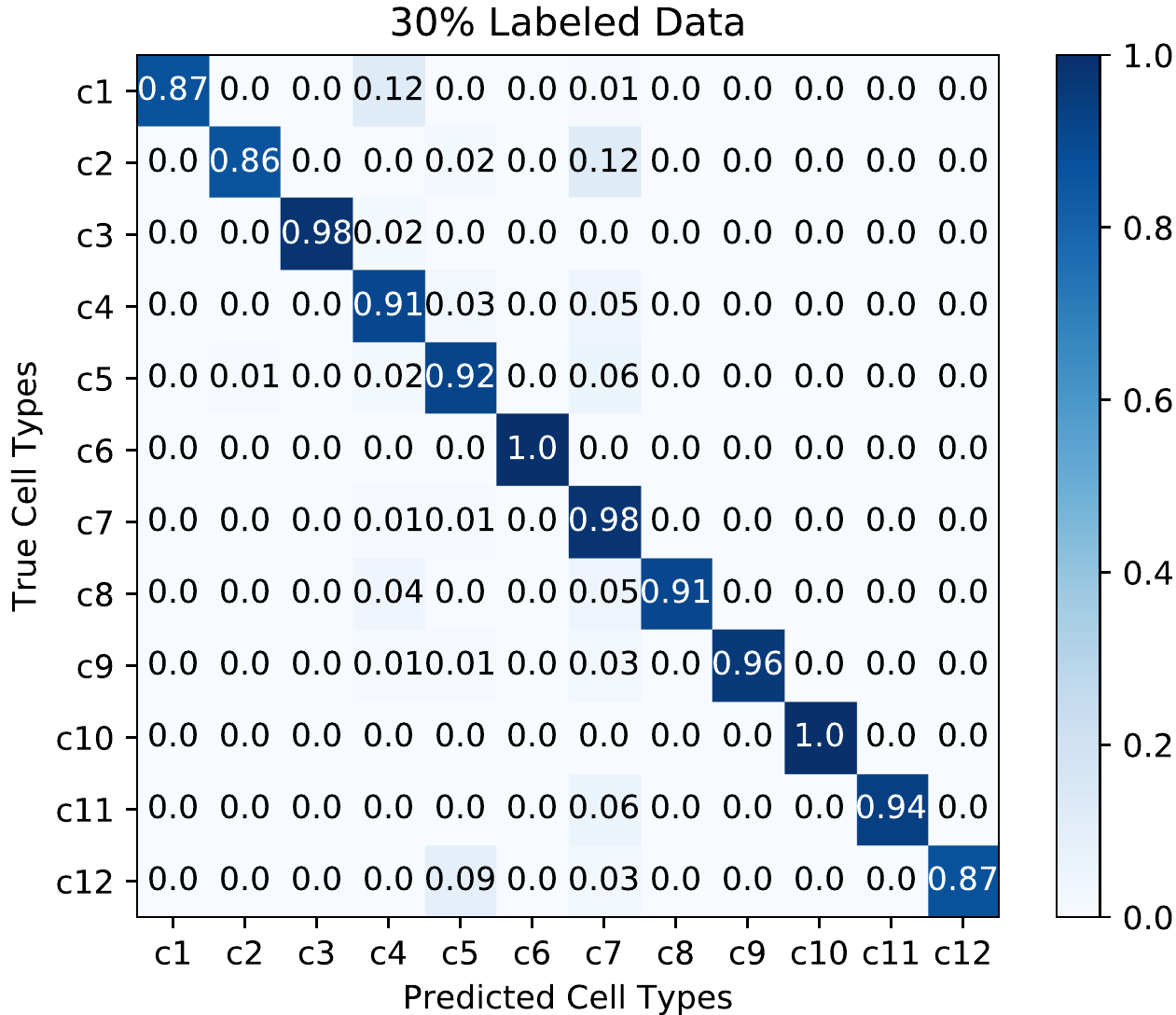} \\
\end{tabular}
 \end{center}
 \caption{Comparison of confusion matrix on different cell types generated with batch size 128 and 256. The left column is for the case of 128 while the right column is for the case of 256.}
 \label{Fig4_cell}
\end{figure*}

Moreover, we compare the confusion matrix for two cases of batch sizes to check the effects with different hyper-parameters in detail, which is shown in Figure \ref{Fig4_cell}. To sum up, for the majority cell type $c_{6}$, the performance is enhanced for the case of larger batch size. For the minority cell types, when employing larger batch size to build the model, the performances for some cell types such as $c_{1}$ and $c_{2}$ are decreased whereas for the cell types like $c_{9}$ and $c_{11}$, the accuracy is increased. It means that we have to choose the optimal batch size for improving the performance of certain minority cell types.

On the other hand, compared to the case with more labeled data, the case with low ratios of labeled cells needs larger batch size to improve the performance for the majority cell type such as $c_{6}$. For instance, when we compare the confusion matrix for the case of 1\% labeled cells, the confusion matrix with batch size 256 has better performance compare to that of  batch size 128. It is consistent to the intuition that with larger batch size, we will obtain larger size of labeled samples to enhance the performance of supervised path when using extremely low ratio of labeled cells. In other words, to improve the performance for the proposed model in the case of extremely  low ratios of labeled data, we should apply larger batch size for the case of majority cell type.

\section{Related Work}
\label{sec5}

Single-cell RNA-seq (scRNAseq) data is able to profile the gene expression levels of cells and to link the dynamics at the molecular level and the cellular level. Analyzing scRNAseq data will be beneficial for obtaining knowledge on cancer drug resistance, gene regulation in embryonic development, and mechanisms of stem cell differentiation and reprogramming~\cite{tang2011development}. In recent years, a lot of progresses have been made on applying bioinformatics techniques to scRNAseq data. However, there still exist many challenges due to dropout events, batch effect, noise, high dimensionality, and scalability~\cite{zheng2019emerging}. 

To overcome these challenges, deep learning techniques have been employed to build effective and efficient computational methods for scRNAseq data. For example, Shaham \textit{et al.} proposed MMD-ResNet to remove batch effect on both mass cytometry and scRNAseq data by combining residual neural networks (ResNets) with the maximum mean discrepancy (MMD)~\cite{shaham2017removal}. To reduce the computational cost, Li \textit{et al.} implemented batch effect removal and clustering in one step~\cite{li2019deep}. Specifically, they built a stacked autoencoder~\cite{hinton2006reducing} to enhance clustering performance. On the other hand, to remove fake zeros, autoencoder based methods such as ``AutoImpute"~\cite{talwar2018autoimpute} and ``DCA"~\cite{eraslan2019single} have been proposed to implement imputation and denoising to address the issue of dropout. Moreover, autoencoder techniques such as denoising autoencoder (DAE)~\cite{lin2017using} and variational autoencoder (VAE)~\cite{ding2018interpretable} have also been applied to reduce dimensions of scRNAseq data~\cite{shaham2017removal, lin2017using, cho2018generalizable}.  In addition, Lopez \textit{et al.} developed an integrative pipeline called “scVI” (single-cell variational inference) to implement multiple tasks including correcting batch effect, removing dropout, imputation, dimension reduction, clustering, and visualization~\cite{lopez2018deep}.  

Recently, Lieberman \textit{et al.} employed transfer learning~\cite{pan2009survey} to reuse a classification scheme that was learned from previous similar experiments for cell type classification~\cite{lieberman2018castle}. However, it is challenging to interpret how transfer learning improve the identification performance in this case. There are several recent works on cell type identification using machine learning techniques such as~\cite{ma2019actinn, lieberman2018castle}. However, these works rely on fully labeled cells to build the identification models, which could not be applied when there are a large number of unlabeled data.


\section{Conclusion and Future Work}
\label{sec7}

In this paper, a novel framework of deep semi-supervised learning is proposed for cell type identification on scRNAseq data. 
As an emerging research area, implementing cell type identification automatically is extremely important for the downstream analysis on the scRNAseq data. However, current methods using supervised learning rely on the availability of large amount of labeled cells, which may not be available in practice. Hence, 
we propose a deep semi-supervised learning model based on recurrent convolutional neural networks (RCNN) that can utilize unlabeled cells to enhance identification performance.  There are two paths in the model for obtaining supervised cross entropy loss  and unsupervised mean squared error loss, respectively. Then training is performed by jointly optimizing these two losses, and this allows the proposed scheme to take advantage of both information from the labeled cells and information from the unlabeled cells. Furthermore, we introduce a preprocessing procedure to overcome the problem of  data sparsity. Experimental results indicate that the proposed model could identify cell type effectively using very limited labeled cells and a large amount of unlabeled cells. In our future work, we plan to extend the proposed model for other tasks such as pathway network construction.

\section*{Acknowledgment}
\label{acknowledgement}
This research work is supported in part by the Texas A\&M Chancellor's Research Initiative (CRI), the U.S. National Science Foundation (NSF) award 1464387 and 1736196, and by the U.S. Office of the Under Secretary of Defense for Research and  Engineering (OUSD(R\&E)) under agreement number FA8750-15-2-0119. The U.S. Government is authorized to reproduce and distribute reprints for Governmental purposes notwithstanding any copyright notation thereon. The views and conclusions contained herein are those of the authors and should not be interpreted as necessarily representing the official policies or endorsements, either expressed or implied, of the U.S. National Science Foundation (NSF) or the U.S. Office of the Under Secretary of Defense for Research and Engineering (OUSD(R\&E)) or the U.S. Government.

\ifCLASSOPTIONcaptionsoff
  \newpage
\fi


\bibliographystyle{IEEEtran}
\bibliography{References}
\end{document}